\documentclass[10pt,letterpaper]{article}
\usepackage[top=0.85in,left=2.75in,footskip=0.75in]{geometry}

\usepackage{amsmath,amssymb}

\usepackage{changepage}

\usepackage[utf8x]{inputenc}

\usepackage{textcomp,marvosym}

\usepackage{cite}

\usepackage{nameref,hyperref}

\usepackage[right]{lineno}

\usepackage{microtype}
\DisableLigatures[f]{encoding = *, family = * }

\usepackage[table]{xcolor}

\usepackage{array}

\newcolumntype{+}{!{\vrule width 2pt}}

\newlength\savedwidth

\newcommand\thickhline{\noalign{\global\savedwidth\arrayrulewidth\global\arrayrulewidth 2pt}%
\hline
\noalign{\global\arrayrulewidth\savedwidth}}

\raggedright
\usepackage[aboveskip=1pt,labelfont=bf,labelsep=period,justification=raggedright,singlelinecheck=off]{caption}

\makeatletter
\renewcommand{\@biblabel}[1]{\quad#1.}
\makeatother

\usepackage{lastpage,fancyhdr,graphicx}
\usepackage{epstopdf}
\fancyheadoffset[L]{2.25in}
\fancyfootoffset[L]{2.25in}
\usepackage{subcaption}

\begin{document}
\vspace*{0.2in}

\begin{flushleft}
{\Large
\textbf\newline{Breaking the Communities:  Characterizing community changing users using text mining and graph machine learning on Twitter} % Please use "sentence case" for title and headings (capitalize only the first word in a title (or heading), the first word in a subtitle (or subheading), and any proper nouns).
}
\newline
\\
Federico Albanese\textsuperscript{1,2*},
Esteban Feuerstein\textsuperscript{2,3},
Leandro Lombardi\textsuperscript{1},
Pablo Balenzuela\textsuperscript{4,5}
\\
\bigskip
\textbf{1} Instituto de Cálculo, Facultad de Ciencias Exactas y Naturales, CONICET- Universidad de Buenos Aires, Buenos Aires, Argentina
\\
\textbf{2} Instituto de Ciencias de la Computación, CONICET- Universidad de Buenos Aires, Buenos Aires, Argentina
\\
\textbf{3} Departamento de Computación, Facultad de Ciencias Exactas y Naturales, Universidad de Buenos Aires, Buenos Aires, Argentina
\\
\textbf{4} Instituto de Física de Buenos Aires (IFIBA), CONICET, Argentina
\\
\textbf{5} Departamento de Física, Facultad de Ciencias Exactas y Naturales, Universidad de Buenos Aires, Buenos Aires, Argentina
\\
\bigskip

% Insert additional author notes using the symbols described below. Insert symbol callouts after author names as necessary.
% 
% Remove or comment out the author notes below if they aren't used.
%
% Primary Equal Contribution Note
%\Yinyang These authors contributed equally to this work.

% Additional Equal Contribution Note
% Also use this double-dagger symbol for special authorship notes, such as senior authorship.
%\ddag These authors also contributed equally to this work.

%\textcurrency Current Address: Dept/Program/Center, Institution Name, City, State, Country % change symbol to "\textcurrency a" if more than one current address note
% \textcurrency b Insert second current address 
% \textcurrency c Insert third current address


% Group/Consortium Author Note
%\textpilcrow Membership list can be found in the Acknowledgments section.

% Use the asterisk to denote corresponding authorship and provide email address in note below.
* falbanese@dc.uba.ar

\end{flushleft}
\section*{Abstract}

Even though the Internet and social media have increased the amount of news and information people can consume, most users are only exposed to content that reinforces their positions and isolates them from other ideological communities. This environment has real consequences with great impact on our lives like severe political polarization, easy spread of fake news, political extremism, hate groups and the lack of enriching debates, among others. Therefore, encouraging conversations between different groups of users and breaking the closed community is of importance for healthy societies. In this paper, we characterize and study users who break their community on Twitter using natural language processing techniques and graph machine learning algorithms. In particular, we collected $9$ million Twitter messages from $1.5$ million users and constructed the retweet networks. We identified their communities and topics of discussion associated to them. With this data, we present a machine learning framework for social media users classification which detects ``community breakers", i.e. users that swing from their closed community to another one. 
A feature importance analysis in three Twitter polarized political datasets showed that these users have low values of PageRank, suggesting that changes are driven because their messages have no response in their communities. This methodology also allowed us to identify their specific topics of interest, providing a fully characterization of this kind of users.

% Please keep the Author Summary between 150 and 200 words
% Use first person. PLOS ONE authors please skip this step. 
% Author Summary not valid for PLOS ONE submissions.   
%\section*{Author summary}

%\linenumbers

% Use "Eq" instead of "Equation" for equation citations.
\section*{Introduction}
Technologically mediated Social networks flourished as a social phenomenon at the beginning of this century with exponents such as Friendster (2002) or Myspace (2003) \cite{island2013friendster} but other popular websites soon took their place. Twitter is an online platform where news or data can reach millions of users in a matter of minutes \cite{lopez2019emotions}.

People with different political opinions and diverse backgrounds interact on this social network. However, this diversity does not translate to enriching debates between users with different profiles because they tend to cluster according to their believes, constituting  homogeneous communities known as echo chambers \cite{jamieson2008echo}. 

Aruguete et al. focused on the interaction between users in political contexts and described how Twitter users frame political events by sharing content exclusively with like-minded users forming two well-defined communities \cite{aruguete2018time}. A segregated partisan structure with extremely limited connection between communities of users with different political orientations on the retweet networks can be found in multiple papers, in different contexts and countries like, for instance, the 2011 parliament elections in Germany \cite{dang2013investigation}, the climate change debate \cite{haussler2018heating}, the \#BlackLivesMatter movement \cite{stewart2018examining}, the 2010 U.S. congressional midterm elections \cite{conover2011political}, the 2011 Canadian Federal Election \cite{gruzd2014investigating}, The Egyptian pro/anti-military intervention debate \cite{borge2015content} or tweets about the death of Venezuelan President Hugo Chavez \cite{morales2015measuring}. The presence of well defined communities can also be found in different platforms and social media \cite{Cinelli2021echo, quattrociocchi2016echo, gilbert2009blogs}.

Most scientific works focused on the dramatic consequences and negative effects of closed communities and echo chambers, which include the increase of negative discourse, hate speech and political extremism \cite{lima2018inside, del2016echo}, confirmation bias (i.e. the users tendency to seek out and receive information that strengthens their preferred narrative) \cite{del2017mapping, quattrociocchi2016echo, gilbert2009blogs, barbera2015tweeting} and spreading of misinformation, baseless rumors and fake news \cite{del2016spreading, choi2020rumor, tornberg2018echo} (one of the main threats to our society according to the World Economic Forum \cite{WEF}).

In this context, the large consumption of information through social networks and its consequences make it essential to analyze the behavior of these communities and think about mechanisms that break them. In this paper, we propose a machine learning framework in order to characterize the {\it community breakers} (i.e. the Twitter users that first belonged to a well defined community and then start interacting mostly with different users swinging to another community). Moreover, once we are able to correctly determine these users, we seek to identify their topics of interest, something that may be useful not only for the sake of understanding but also to intervene in the dynamics of the discussion.
%since they could be useful in order to encourage conversations between different communities and breaking the echo chambers at a bigger scale.

Three datasets were built and used in order to show that the methodology can be easily generalized to different scenarios. Namely, we examined three Twitter network datasets constructed with tweets from: 2017 Argentina parliamentary elections, 2019 Argentina presidential elections and 2020 tweets about Donald Trump.
For each dataset, we analyzed two different time periods and identified the larger communities corresponding to the main political forces. Using graph topological information and detecting topics of discussion of the first network, we built and trained a model that classifies whether an individual will change his/her community, identifying the topics of interest and relevant features of the community breakers.

Our main contributions are the following:

\begin{enumerate}
    %\item We confirmed the presence of echo chambers in these twitter political conversations.
    \item We describe a generalized machine learning framework for social media users classification, in particular, to detect and characterize community changing users. This framework includes natural language processing techniques and graph machine learning algorithms in order to describe the features of each individual. 
    \item We experimentally analyze the machine learning framework by performing a feature importance analysis. While previous works used text, Twitter profiles and some twitting behavior characteristics to automatically classify users with machine learning \cite{conover2011predicting,pennacchiotti2011democrats,benevenuto2010detecting,borondo2012characterizing}, here we show the value of adding graph features in order to identify the label of a user. In particular we ascert the importance of the low value of  ``PageRank''\cite{page1999pagerank} measure for this specific task. A possible interpretation of this result is that a person changes their community because their massage was not heard in their previous community.
    \item We also identify the topics that are considerably more relevant to the community breakers. Identifying these key topics has a valuable impact for social science and politics.
\end{enumerate}

The paper is organized as follows. In the Data Collection section, we describe the data used in the study. In the Methods section, we describe the graph unsupervised learning algorithms and other graph metrics that were used, the natural language processing tools applied to the tweets and a machine learning model for the task of classifying the community breakers. In the Results section, we analyze the classifying model and which are the important characteristics of these users. Finally, we interpret these results in the Conclusions section. 
 %The code is in github (omitted for anonymity reasons).

\section*{Data Collection}
 \label{sec:data}

Twitter has several APIs available for developers. Among them is the Streaming API that allows the developer to download in real time a sample of tweets that are uploaded to the social network filtering it by language, terms, hashtags, etc. \cite{makice2009twitter,morstatter2013sample}. The data is composed of the tweet id, the text, the date and time of the tweet, the user id and username, among other features. In case of a retweet, it has also the information of the original tweet's user account. 

For this research, we collected three datasets in two different periods of time: 2017 Argentina parliamentary elections (2017ARG), 2019 Argentina presidential elections (2019ARG) and 2020 United States tweets of Donald Trump (2020US). For the Argentinian dataset, the Streaming API was used during the week preceding the primary elections and the week before the general elections. Keywords were chosen according the four main political parties present in the elections. Details and context can be found in the supporting information. For the 2020US dataset, we used ``realDonaldTrump'' (the official account of president Donald Trump) as keyword and the weeks from May $9^{th}$ to May $16^{th}$ and from June $10^{th}$ to June $16^{th}$ of 2020 as first and second time period respectively. Twitter messages are in the public domain and only public tweets filtered by the Twitter API were collected for this work. For the purpose of this research, we have analyzed more than 9 million tweets and more than 1.5 million individuals in total. The specific start and end collection date, the total number of tweets and users of each dataset can be seen in Table \ref{table1}.

\begin{table}[!ht]
\begin{adjustwidth}{-2.25in}{0in} % Comment out/remove adjustwidth environment if table fits in text column.
\centering
\caption{
{\bf The total number of tweets and users of each dataset for both time periods.}}
\begin{tabular}{|l|l|l|l|l|l|l|}
\hline
Dataset &\multicolumn{2}{|l|}{\bf 2017ARG} & \multicolumn{2}{|l|}{\bf 2019ARG} & \multicolumn{2}{|l|}{\bf 2020US}\\ \thickhline
Year & \multicolumn{2}{|l|}{2017} & \multicolumn{2}{|l|}{2019} &  \multicolumn{2}{|l|}{2020}\\ \hline
Start date & August $7^{th}$ & October $15^{th}$ & August $5^{th}$ & October $20^{th}$ &  May $9^{th}$ &June $10^{th}$\\ \hline
End date & August $13^{th}$ & October $22^{th}$ & August $12^{th}$ & October $27^{th}$ &  May $16^{th}$ & June $16^{th}$\\ \hline
\# Tweets & 2117708 & 1751813 & 1638585 & 1587563 & 1148032 & 1137785\\ \hline
\# Users & 86361 & 298866 & 256860 & 271910 & 335077 & 348793\\ \hline 
\end{tabular}
\label{table1}
\end{adjustwidth}
\end{table}

\section*{Methods}
\label{sec:methods}

In this section, we will present the methodology employed to characterize the Twitter users. We start with the retweet network and the algorithm to find communities. Then, we introduce the  different metrics which describe the interaction's networks among them. We also extract the text features of the tweets using a natural language processing algorithm. Finally, we describe a supervised learning model which uses the individual's characteristics as instances and predicts which users change their community. These models allow us to highlight which user features characterize the community breaking users. 

\subsection*{The retweet network}
\label{sec:RN}
We represent the interaction among individuals in terms of a graph $G = (N,E)$, where users are nodes ($N$) and retweets between them are edges ($E$). Considering that a user can be retweeted multiple times by another user, this is well modelled by a directed and weighted graph. However, when a user $n_1$ retweets a tweet written by another user $n_2$, should the edge point form $n_1$ to $n_2$  or from $n_2$ to $n_1$? This definition has important implications. 
In the first scenario, the edges represent pointers to the ``influencers'' and important content generators. In the second scenario, the edges represent the flow of information through the network, going from the source to the user who spread the message. 
Indeed, there is no clear consensus in the scientific literature about which direction should be given to the edges: while some authors \cite{amati2016retweet, kogan2015think, yang2012finding} use the first, others \cite{rath2017retweet, morales2015measuring, conover2011political} prefer the second one. 
Although they are symmetrical, they are different, and we cannot tell a priori which one is better for our purpose, so we decided to calculate the topological features in both scenarios. We named the directions of the edges CR (from content Creator to Retweeter) and RC (from Retweeter to content Creator).
Isolated nodes (never retweeting nor retweeted) were not taken into account for this analysis.

\begin{figure}[!h]
  \centering
  \includegraphics[width=\linewidth]{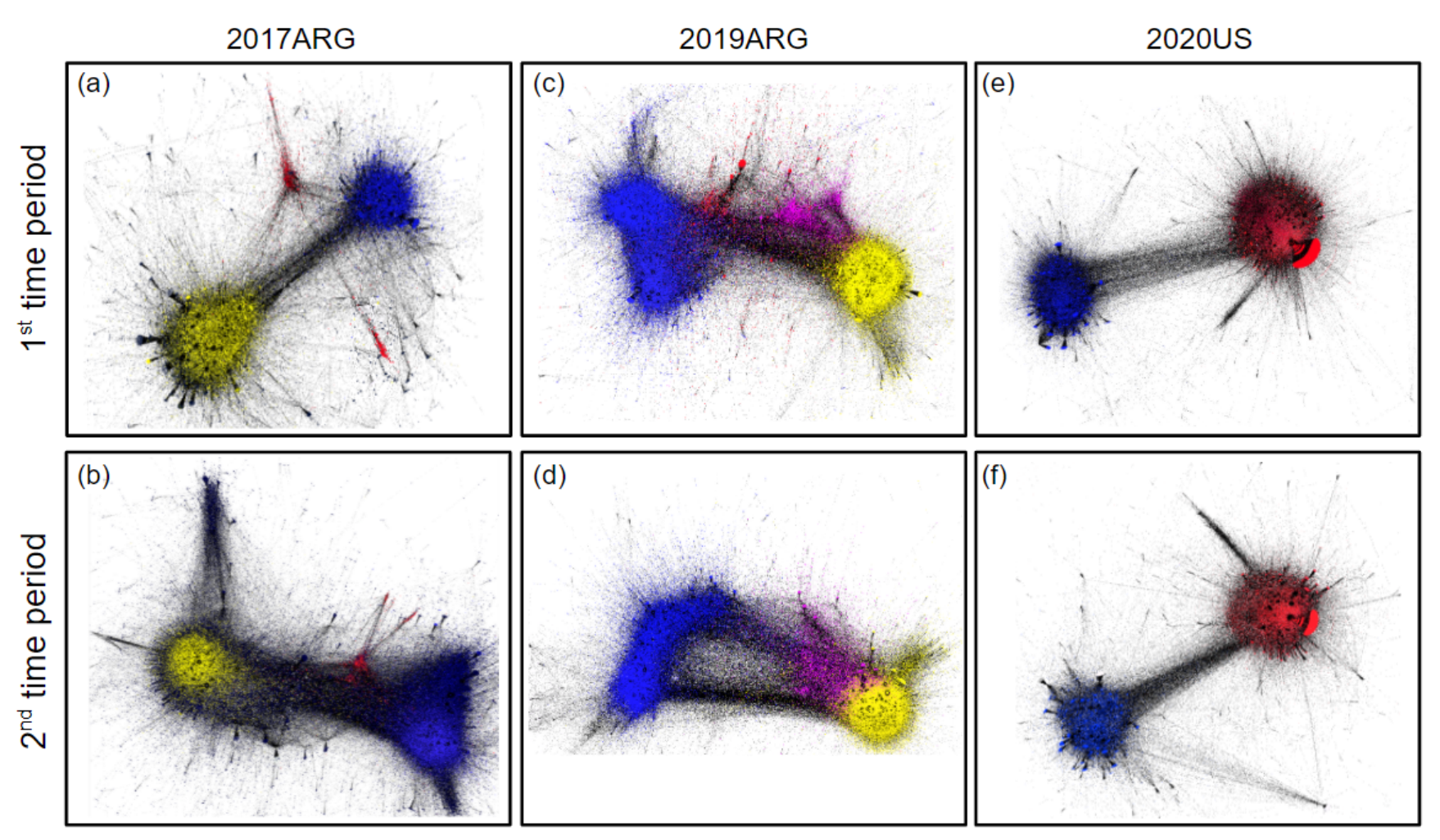}
  \caption{{\bf Retweet networks:} (a) 2017 Argentina primary election; (b) 2017 Argentina general election; (c) 2019 Argentina primary election; (d) 2019 Argentina general; (e) first time period of the 2020US and (f) second time period of the 2020US. Each node is a Twitter user (colored depending on its community) and each edge (directed and weighted) represents the retweets between two given users (in black).}
  %\Description{In the 6 graphs, individuals are grouped in communities. In the cases of Argentina there are two major communities and several minor ones, while in the USA dataset most of the users are grouped into only two communities.}
  \label{gephi1}
\end{figure}

In Fig \ref{gephi1}, we can visualize the retweet network for each time period and dataset. In the case of the US dataset, most of the users are concentrated in two groups, portraying the political polarization in that country. On the other hand, in the Argentinean dataset we can identify two large groups and also some smaller ones. The graph visualizations are produced with Force Atlas 2 layout using Gephi software \cite{jacomy2014forceatlas2}.

\subsection*{Unsupervised Learning: Community Detection}
\label{sec:CD}

In a given graph, a community is a  set of nodes  strongly  connected among them and with little or no connection with nodes of other communities \cite{yang2016comparative}. We implement an algorithm to detect communities in large networks which allows us to characterize the users by their relationship with other users. In this context, the {\em modularity} is defined as the fraction of the edges that fall within a given community minus the expected fraction if edges were distributed at random \cite{brandes2007modularity}. The Louvain method for community detection \cite{blondel2008fast} seeks to maximize modularity by using a greedy optimization algorithm. This method was chosen to perform the analysis due to the characteristics of the database. While other algorithms such as Label Propagation are good for large data networks, their performance decreases if clusters are not well defined \cite{lancichinetti2009community}. In contrast, in these cases the Louvain or Infomap methods obtain better results. However, for the size of our graphs (in the order of hundreds of thousands of nodes and about one million edges), the Louvain method is more efficient than the other ones in terms of computation time required because it scales roughly linearly with the number of edges ~\cite{emmons2016analysis}.

Despite having found several communities, we just considered the largest ones for each case. For the 2017ARG and 2019ARG datasets we used the four biggest communities because, when examining the text of the tweets and the users with the highest degree, each one had a clear political orientation corresponding to the four biggest political parties in the election. These communities are labeled as ``Cambiemos'', ``Unidad Ciudadana'', ``Partido Justicialista'' and ``1 Pais'' for 2017ARG and ``Frente de Todos'', ``Juntos por el Cambio'', ``Consenso Federal'' and ``Frente de Izquierda-Unidad'' for 2019ARG (electoral context is provided in the supporting information). 
Regarding the 2020US dataset, we used the 2 biggest communities because of the bipartisan political system of the United States (Republicans and Democrats) and the clear structure present in the retweet networks, where only two big clusters concentrate almost all of the users and interactions (see Fig \ref{gephi1}). In contrast, the Argentinean election datasets have two principal communities and some minor communities as well. This network topology with highly connected and polarized clusters had been reported in previous works \cite{garimella2018quantifying, aruguete2018time, conover2011political, stewart2018examining, brady2017emotion, cherepnalkoski2015retweet}.

Given the stochasticity of the method, we follow the solution proposed by Lancichinetti et al. \cite{lancichinetti2012consensus} that runs the Louvain method several times ($100$ in our case) and, in order to minimize the possibility of an incorrect labeling, keeps for the machine learning task only the nodes that were always consistently assigned to the same community in all iterations. 

\subsection*{Graph Features}
\label{sec:GF}

Given that the analyzed datasets comprise two snapshots of the retweet network separate in time, we need to fully characterize the users in the early networks in order to properly identify those users that change their community. With this goal, we computed the following metrics for each user in the network: Degree, Indegree, Outdegree, PageRank \cite{page1999pagerank}, betweenness centrality \cite{brandes2001faster}, clustering coefficient \cite{saramaki2007generalizations} and cluster affiliation (the community detected by the Louvain method).

As we mentioned earlier, it's important to note that the direction of the edges of the network drastically affects the value of these metrics. Consequently, we calculated them with both interpretations. 

All these metrics were used as features in the machine learning classification task and feature importance analysis.

\subsection*{Natural Language Processing Features}
\label{sec:NLP}

In order to determine the topics of discussion during the first period of each dataset, we analyzed the text of the tweets using natural language processing analysis and we calculated a low dimensional embedding for each user.

The tweets were described as vectors through the Term Frequency - Inverse Document Frequency (tf-idf) representation \cite{ramos2003using}. Each value in the vector corresponded to the frequency of a word in the tweet (the term frequency, \textit{tf}) weighted by a factor which measures the degree of specificity (inverse document frequency, \textit{idf}). We used 3-grams and a modified stop-words dictionary that not only contained articles, prepositions, pronouns and some verbs but also the names of the politicians, parties and words like ``election''.

Then, we constructed a matrix $M$ concatenating the tf-idf vectors, with dimensions the number of tweets times the number of terms. We performed topic decomposition using Non-Negative Matrix Factorization (NMF) \cite{xu2003document} on the matrix $M$. NMF is an unsupervised topic model which factorizes the matrix $M$ into two matrices $H$ and $W$ with the property that all three matrices have no negative elements. We selected the NMF algorithm because this non-negativity makes the resulting matrices easier to inspect and to understand their meaning. The matrix $H$ has a representation of the tweets in the topic space, in which the columns are the degree of membership of each tweet to a given topic. On the other hand, the matrix $W$ provide the combination of terms which describes each topic \cite{Albanese_2020}. 

The obtained results, analyzing just the tweets corresponding to the first time period, are detailed in the supporting information. The decomposition dimension was swept between 5 and 30, and for each dataset we chose a number of topics in the corpus so as to have a clear interpretation of each one. The same methodology was used and described in \cite{pinto2019quantifying, Albanese_2020}.

Once we collected all this information, Twitter users were also characterized by a vector of features where each cell corresponds to one of the topics and its value to the percentage of tweets the user tweeted with that topic.

\subsection*{Feature importance analysis}
\label{sec:MLcm}

Given that our objective was to characterize users who ``break'' their community and start interacting with users from other clusters, we implemented a machine learning model which classifies users and then performed a feature importance analysis. 
The instances of the model were the Twitter users who were active during both time periods \cite{cazabet2019challenges} and belonged to one of the biggest communities in both time periods networks. Consequently, the number of users considered at this stage was reduced. Individuals were characterized by a feature vector with components corresponding to the mentioned topological metrics and others corresponding to the percentage of tweets in each one of the topics of interest extracted with Non-negative matrix factorization. The information used to construct these embedding was gathered from the whole first time period retweet network. The target was a binary vector that takes the value $1$ if the user changed communities between the first and the second time periods (a community breaker) and $0$ otherwise (not a community breaker). 
The summary of the datasets is shown in Table \ref{table2}. 

\begin{table}[!ht]
\begin{adjustwidth}{-2.25in}{0in} % Comment out/remove adjustwidth environment if table fits in text column.
\centering
\caption{
{\bf Summary of the datasets used in the experiments.}}
\begin{tabular}{|l|l|l|l|}
\hline
 & {\bf 2017ARG} & {\bf 2019ARG} & {\bf 2020US}\\ \thickhline
$\# Individuals$ & 21134 & 26118 & 116854 \\ \hline
$\# Communities$ & 4 & 4 & 2 \\ \hline
$\# Text Features$ & 9 & 7 & 6 \\ \hline
$\# Graph Features$ & 10 & 10 & 10 \\ \hline
$Training set size$ & 14159 & 17499 & 78292 \\ \hline
$Test set size$ & 6975 & 8619 & 38562 \\ \hline
\end{tabular}
\label{table2}
\end{adjustwidth}
\end{table}

The gradient boosting technique uses an ensemble of predictive models to perform the task of supervised classification and regression \cite{chen2016xgboost}. These predictive models are then optimized, iteration by iteration, using the gradient of the cost function of the previous iteration. In this scenario, XGBoost, a particular implementation of this technique, had proven to be efficient in a wide variety of supervised scenarios outperforming previous models \cite{nielsen2016tree}.

We used a $67/33$ random split between train and test. In order to do hyper-parameter tuning of the XGBoost models, we used the randomized search method \cite{bergstra2012random} over the training dataset with 3-fold cross-validation, which consists of trying different random combinations of parameters to find an optimum.

Finally, we performed random permutation of the features values among users in order to understand which of them are the most important in the performance of our model (using the so-called Permutation Feature Importance algorithm \cite{altmann2010permutation}). In these way, we could identify the most important characteristics that separates the users that do change their community form those that do not change who they interact with.

\section*{Results}
\label{sec:results}

\subsection*{Topics of interest}
\label{sec:PersuasiveTopics}
%might be more important to
Users tweet about different topics. Some discussed topics are more frequent in users that change their community than in the general audience. Considering that most users do not change their community and always interact with the same users, a simple analysis of the hole dataset and listing Twitter trending topics may not be a good representation of their interest. Consequently, we defined a ``community breaking topic'' as a topic used primarily by the community breakers and not used by other users.

With the intention of doing a deeper analysis of the topic embedding for each dataset, we first enumerate the main topics in each corpus. For the 2020US dataset:

\begin{enumerate}
\item President Donald Trump: The $45^o$ President of the United States.
\item Obamagate: The accusation that Barack Obama is conspiring against Donald Trump.
\item World Health Organization: President Trump announcing the US will pull out of the World Health Organization.
\item Thank you: Individuals thanking President Trump for his policies in regard to the COVID-19 pandemic.
\item Fake news: Individuals discussing and claiming that certain news are fake.
\item President Barack Obama: The $44^o$ President of the United States and his administration.
\end{enumerate}

\begin{figure}[!h]
  \centering
  \includegraphics[width=\linewidth]{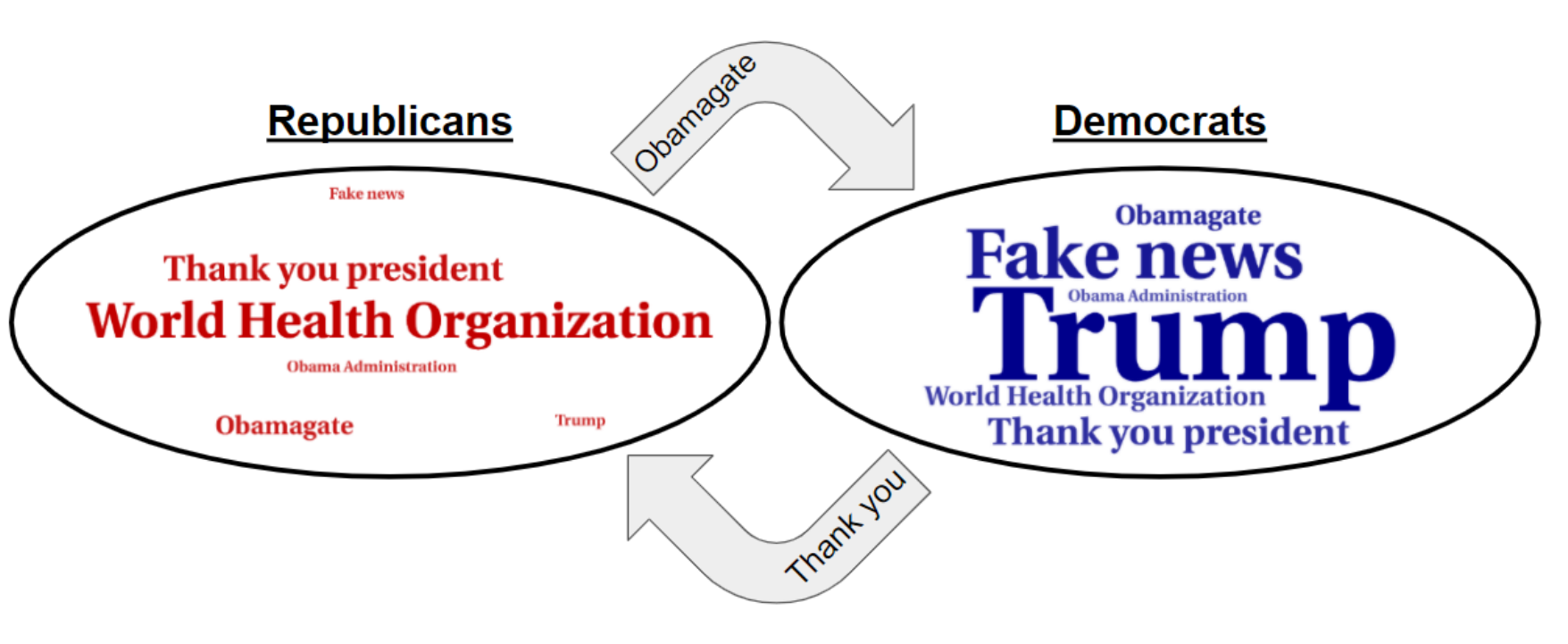}
  \caption{{\bf Community topics:} The sizes of the word clouds represent the importance of a topic in each community. the arrows indicate the flow of users between the communities of the primary elections to the general elections. The topics on the arrows show the most frequent topic among the users that change between those communities.}
  \label{Topic2020US}
\end{figure}

The two most used topics found were the U.S. president ``Donald Trump'' and ``the World Health Organization''. However, Fig \ref{Topic2020US} shows that these topics were primarily used by users that did not change their community. In contrast, the ``Obamagate'' topic was used by users that change from the republican community to the democrat community. On the other hand, the topic ``Thank you'', where people thanks and vindicates president Donald Trump health policies, was the main topic used by users that change from the Democrat community to the Republican one. Considering that these last two topics were more used by community breakers than others, we refer them as ``community breaking topics''. In contrast, other topics such as ``World Health Organization'' or ``Donald Trump'' were commonly used by most of the users but not by the users who altered the users they interact with.

Equivalent analysis can be done with the other two corpora. For the 2017ARG dataset the main topics were the following:

\begin{enumerate}
\item Santiago Maldonado: An activist who disappeared on August 1, 2017, after a gendarmerie repression of a riot.
\item Missing ballots: People reporting missing ballots on Twitter.
\item Election official and poll workers: People tweeting about the absence of election officials when voting in some schools.
\item Economy: This topic particularly focused on poverty, unemployment and structural adjustment programs.
\item ``Once" Tragedy: The 2012 rail disaster in which 51 persons died and 703 were injured during the presidency of Cristina Kirchner in Buenos Aires' train station ``Once".
\item Exit polls: The results of opinion polls of people leaving the polling station.
\item Santa Cruz: An important Argentinean province whose Governor, Alicia Kirchner, belongs to the Opposition. 
\item Buenos Aires: The biggest and most populated province in Argentina.
\item Venezuela: Tweets about the government of Venezuela and its relationship with Cristina Kirchner.
\end{enumerate}

\begin{figure}[!h]
  \centering
  \includegraphics[width=\linewidth]{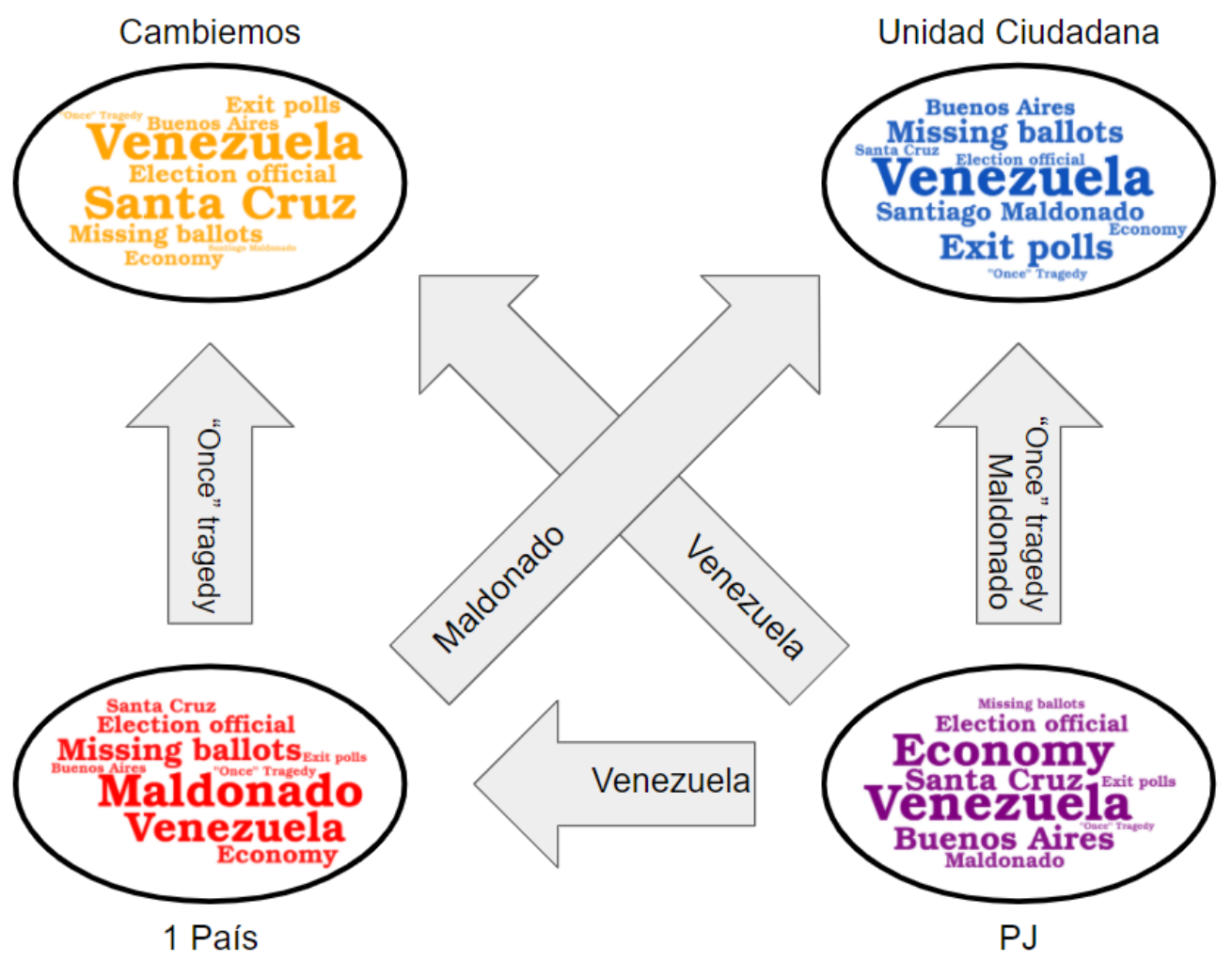}
  \caption{{\bf Community topics:} The sizes of the word clouds represent the importance of a topic in each community. the arrows indicate the flow of users between the communities of the primary elections to the general elections. The topics on the arrows show the most frequent topic among the users that change between those communities. When the percentage of users that changed from one community to the other was less than $ 1 \% $, the corresponding arrow is not drawn.}
  \label{Topic2017ARG}
\end{figure}

In Fig \ref{Topic2017ARG}, the most important topics for the 2017ARG classifier are ``Economy'', ``Once Tragedy''  and ``Santiago Maldonado''. We can contextualize these results by looking which are the main topics discussed in each community as well the ones discussed among the users that change between them. We can see that  ``Venezuela''  is one of the most discussed topics in the people remaining in four communities and ``Santiago Maldonado'' is a relevant topic in the communities ``Unidad Ciudadana'' and ``1 Pais''. When we look at the main topics discussed by users that change their communities between elections, we can observe that ``Venezuela'' identifies those that go from ``Partido Justicialista (PJ)'' to ``1 Pais'' and ``Cambiemos'' meanwhile ``Santiago Maldonado'' is a key topic among those who arrive to ``Unidad Ciudadana'' from ``Partido Justicialista (PJ)'' and ``1 Pais''. The topic ``Once Tragedy'' is primeraly used by the users that change from ``1 Pais'' to ``Cambiemos''. Considering that these topics are considerably more used by the users who change their community than by the other users, it can be affirmed that these are ``community breaking topics''. In contrast, other topics such as ``Economy'' or ``Santa Cruz'' were also commonly used by most of the users but not by the users who altered the users they interact with.

%with non-negative matrix factorization
Lastly, the topic embedding obtained for the 2019ARG dataset  are the following: 

\begin{enumerate}
\item Election acts
\item the Government of the Ciudad Autonoma de Buenos Aires: the capital city of Argentina.
\item The Argentinean province of Buenos Aires: The biggest and most populated province in Argentina.
\item The Argentinean cities of Cordoba and Rosario: Two of the  most populated cities in Argentina.
\item Cesar Milani: Former Lieutenant general during the administration of Cristina Kirchner.
\item Maria Eugenia Vidal: Former Governor of the Buenos Aires province and candidate of ``Juntos por el Cambio''.
\item Economy: This topic particularly focused on poverty and unemployment.
\end{enumerate}

\begin{figure}[!h]
  \centering
  \includegraphics[width=\linewidth]{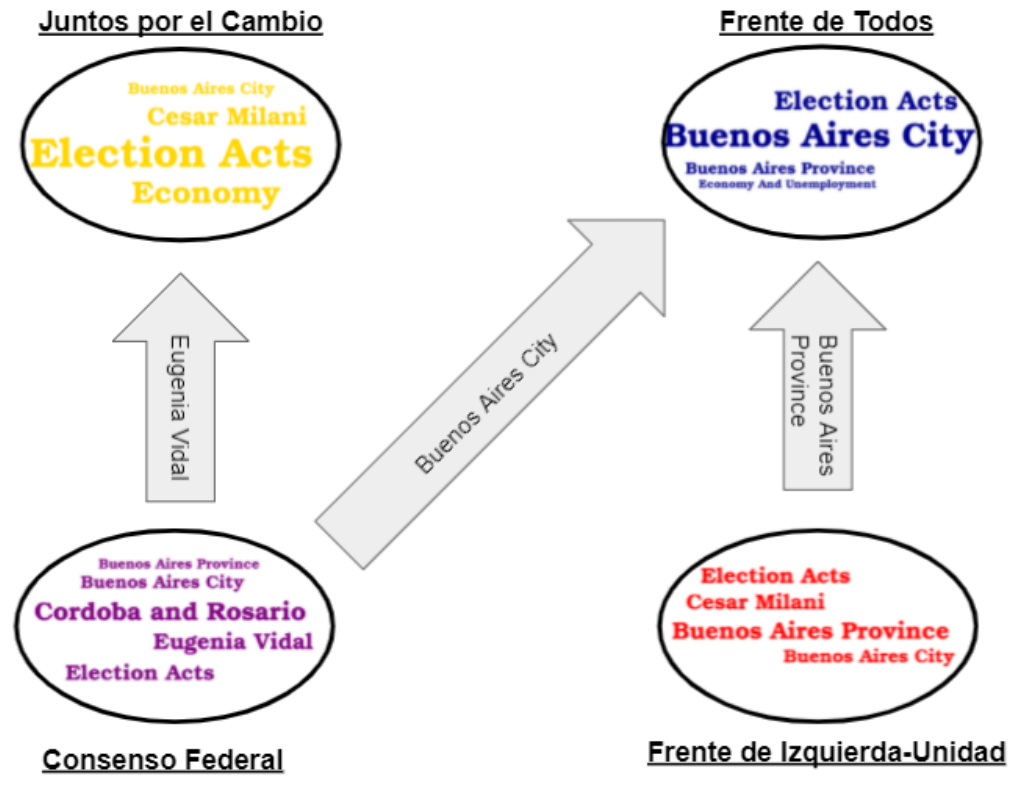}
  \caption{{\bf Community topics:} The sizes of the word clouds represent the importance of a topic in each community. the arrows indicate the flow of users between the communities of the primary elections to the general elections. The topics on the arrows show the most frequent topic among the users that change between those communities. When the percentage of users that changed from one community to the other was less than $ 1 \% $, the corresponding arrow is not drawn.}
  \label{Topic2019ARG}
\end{figure}

The Election acts with the final results was one of the major topic in all of the four biggest communities, as it can be seen in Fig \ref{Topic2019ARG}. In these figure it can also be observed which were the most important topics of each community and wich were the topics of interest of the users that change their community.

\subsection*{Classification model}
\label{sec:predict}

We trained three different gradient boosting models for each dataset: the first one was trained only with the features obtained via text mining (how many tweets of the selected topics the user talks about); a second one was trained just with features obtained through complex network analysis (degree, PageRank, betweeness centrality, clustering coefficient and cluster affiliation); and the last one  was trained with all the data. In this way, we could compare the importance of the natural language processing and the complex network analysis for the task of classifying community changing users.

In table \ref{TableAUC} we can see the area under the ROC (receiver operating characteristic) curve \cite{rice2005comparing} of the different models for each dataset. The best performance is obtained in all cases by the machine learning model built with all the features of the users, which is able to more efficiently predict the community breakers. This result is expected, since an assembly of models manages to have sufficient depth and robustness to understand the network information, the topics of the tweets and the graph characteristics of the users.

\begin{table}[!ht]
\begin{adjustwidth}{-2.25in}{0in} % Comment out/remove adjustwidth environment if table fits in text column.
\centering
\caption{
{\bf Summary of the results of the XGBoost models (ROC AUC). }}
\begin{tabular}{|l|l|l|l|}
\hline
 & {\bf 2017ARG} & {\bf 2019ARG} & {\bf 2020US}\\ \thickhline
XGBoost (text features) & 0.7339 & 0.6683 & 0.6839 \\ \hline
XGBoost (graph features) & 0.7664 & 0.7995 & 0.7425 \\ \hline
{\bf XGBoost (text + graph features)} & {\bf 0.7925} & {\bf 0.8019} & {\bf 0.7614} \\ \hline
\end{tabular}
\label{TableAUC}
\end{adjustwidth}
\end{table}
%better defining characteristics
In table \ref{TableAUC} we can also observe that the graph features are most informative features than the text ones in order to classify users, since the model with only graph features has a higher score than the model with only text features.

\begin{figure}[!h] 
  \centering
  \begin{subfigure}{0.6\columnwidth}
         \captionsetup{justification=centering}
         \centering
         \includegraphics[width=\textwidth]{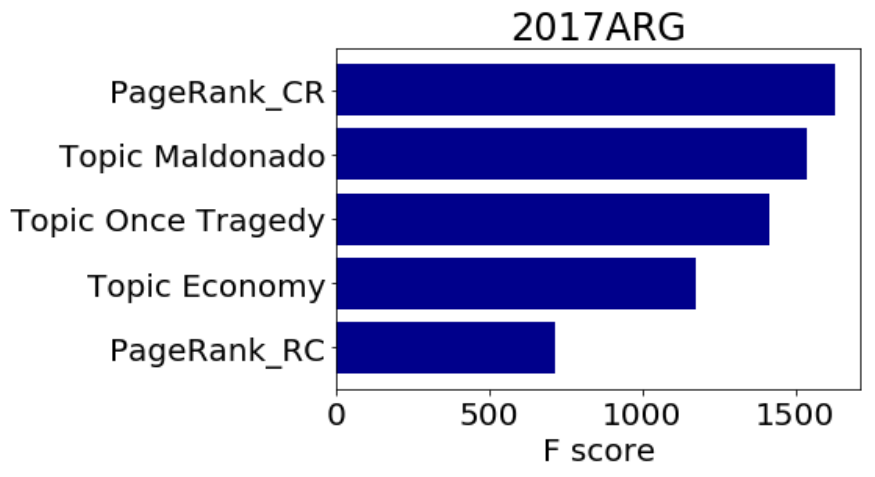}
         \caption{2017ARG}
  \end{subfigure}
  \hfill
  \begin{subfigure}{.6\columnwidth}
         \captionsetup{justification=centering}
         \centering
         \includegraphics[width=\textwidth]{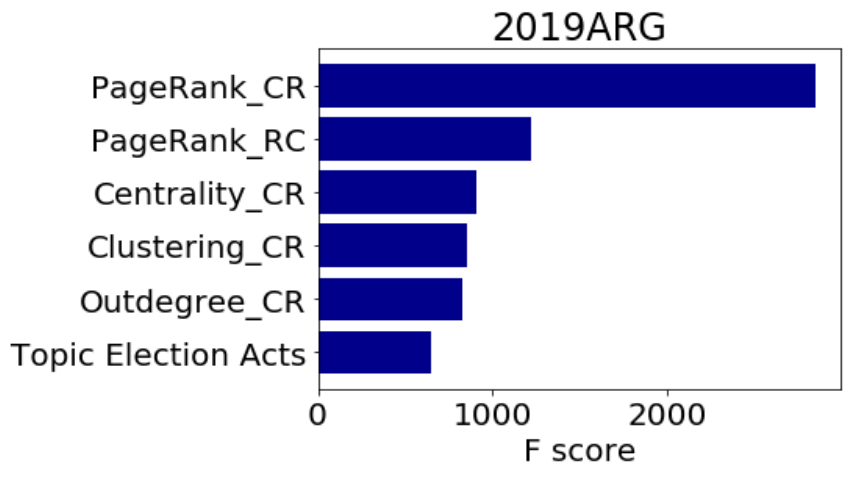}
         \caption{2019ARG}
  \end{subfigure}
  \hfill
  \begin{subfigure}{.6\columnwidth}
         \captionsetup{justification=centering}
         \centering
         \includegraphics[width=\textwidth]{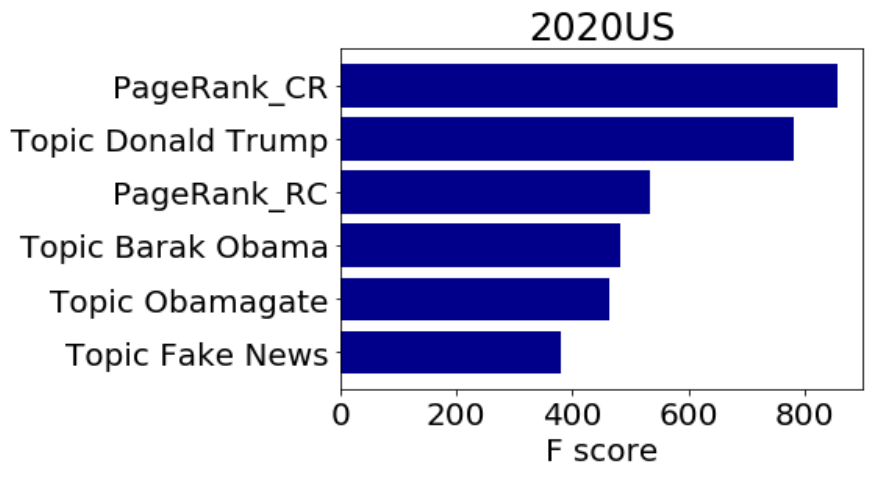}
         \caption{2020US}
  \end{subfigure}
  \hfill
  \caption{{\bf Feature importance analysis:} The most important features according to the results of the permutation feature importance of the XGBoost model, for the dataset 2017Arg, 2019Arg and 2020US (from top to bottom respectively). CR (Content Creator $\rightarrow$ Retweeter) and RC (Retweeter $\rightarrow$ Content Creator) indicates the direction of the edges with which that feature was calculated.}
  %\Description{The ROC curves show how in the three cases the XGBoost model trained with all the features obtains the best performance, followed by the XGBoost model trained only with graph features and the XGBoost model with text features. The  baselines have the worst performance.}
  \label{PFI}
\end{figure}

We performed random permutation of the features values among users in order to understand which of them are the most important in the performance of our model (using the so-called Permutation Feature Importance algorithm \cite{altmann2010permutation}). In Fig~\ref{PFI}, we observed that the most important feature in all cases corresponds to the node's connectivity: $PageRank_{CR}$, where the edges point from the tweet source (the content creator) to the user who retweeted. In contrast, the other $PageRank_{RC}$ (corresponding to the other direction of the edges), had a lower importance feature coefficient in all three datasets. These means that there is a clear privileged direction of edges for the task of detecting the the community breakers. 

When comparing the $PageRank_{CR}$ (PR) averages of the community breakers  with the other users, we observed that the latter had higher values in all cases (Table \ref{TablePagerank}). We applied the  Kolmogorov-Smirnov test \cite{hodges1958significance} to the PR distributions of each set and found that these differences were statistically significant in all cases ($p < 0.001$).

\begin{table}[!ht]
\begin{adjustwidth}{-2.25in}{0in} % Comment out/remove adjustwidth environment if table fits in text column.
\centering
\caption{
{\bf $PageRank_{CR}$ (PR) average.}}
\begin{tabular}{|l|l|l|l|}
\hline
 & {\bf 2017ARG} & {\bf 2019ARG} & {\bf 2020US}\\ \thickhline
PR of community breakers & 1.32e-5 & 3.81e-6 & 3.16e-6 \\ \hline
PR of non community breakers & 1.55e-5 & 4.43e-6 & 3.47e-6 \\ \hline
\end{tabular}
\label{TablePagerank}
\end{adjustwidth}
\end{table}

The $Pagerank$ measures how relevant or important a user is in the retweet network based on the retweets of their messages and the importance of the users who retweeted. The direction of $PageRank_{CR}$ represents the information flow in a network, starting from the tweet creator and then spreading throw the network. The fact that the community breakers had statistically lower $PageRank_{CR}$ values means that these users were less relevant to the tweeter conversation and their messages did not spread in their original community. A possible interpretation of these results is that a user changes community when their messages have no response by their original community of belonging. 
%This result has a relevant sociological meaning.

The fact that the $PageRank_{CR}$ is the most important feature is also consistent with the model trained with network features getting a better $ AUC $ than the model trained with the texts of the tweets in the three datasets. 

Previous works used text, Twitter profile and some twitting behavior characteristics to automatically classify users with machine learning, but none of them incorporated the use of these graph metrics \cite{conover2011predicting,pennacchiotti2011democrats,benevenuto2010detecting,borondo2012characterizing,tinati2012identifying}. Our work shows the importance of also including these graph features in order to identify the community breakers. 

\section*{Conclusion}
\label{sec:conclusions}

In this paper we presented a machine learning framework approach in order to identify and characterize users who break their community. The framework includes natural language processing techniques to detect their topics of interest and graph machine learning algorithms in order to describe how an individual interacts with other users.

Three datasets were used in this analysis: 2017ARG, 2019ARG and 2020US. These datasets were constructed covering different scenarios: with tweets from two countries, during different political contexts (a parliamentary election, a presidential election and a non-election period) and party system (a multi-party system and a two-party system). The machine learning framework was applied to these different datasets with similar results, showing that the methodology can be easily generalized.

We found that the community breakers had statistically lower values of $PageRank_{CR}$. This graph feature was also the most important indicator of the classification task in all three datasets according to the feature importance analysis. Therefore, our results indicate that this proposed feature does a good job characterizing if a user is a community breaker. This feature was neglected in previous works of user classification on Twitter \cite{conover2011predicting,pennacchiotti2011democrats,benevenuto2010detecting,borondo2012characterizing,tinati2012identifying}. In particular, our results also show that there is a clearly privileged direction on the network for this task, with the edges going from the content creator to the retweeter. A possible interpretation for these last two results is that users change who they interact with when their messages have no response and are not being ``heard" by their community.

Finally, the proposed framework also identifies which of the topics are of interest for these users. Being able to identify the topics that encourage users to interact with other users outside their community is of vital importance in order to reduce the effects of echo chambers such as political extremism, hate groups or the spread of fake news and stimulate enriching debates.
Also, this methodology might be useful for a political party and help them decide which issues should be prioritized in its agenda with the intention of maximizing the number of individuals that migrate to their community. 

Understanding the characteristics and the topics of interest of the community breakers in a polarized environment can provide an enormous benefit for social scientists and political parties. This research intends to supply them with tools to improve their understanding of their behavior. 

\section*{Supporting information}

% Include only the SI item label in the paragraph heading. Use the \nameref{label} command to cite SI items in the text.

\paragraph*{Supporting\_information.pdf}
\label{appendix}
{\bf Extra information:} We described the political context of the datasets and specified the keywords which were used for collecting the tweets using the public Twitter API.

\section*{Acknowledgments}
We thank Sebastián Pinto for careful reading of the manuscript and helpful comments.

\nolinenumbers

% Either type in your references using
% \begin{thebibliography}{}
% Text
% \end{thebibliography}
%
% or
%
% Compile your BiBTeX database using our plos2015.bst
% style file and paste the contents of your .bbl file
% here. See http://journals.plos.org/plosone/s/latex for 

\end{document}